A cryostatic, fast scanning, wideband NQR spectrometer for the VHF range

Hermann Scharfetter, Markus Bödenler, Dominik Narnhofer

Institute of Medical Engineering, Stremayrgasse 16, Graz University of Technology, 8010 Graz

Corresponding Author: Hermann Scharfetter, hermann.scharfetter@tugraz.at

## Abstract

In the search for a novel MRI contrast agent which relies on $T_1$ shortening due to quadrupolar interaction between Bi nuclei and protons, a fast scanning wideband system for zero-field nuclear quadrupole resonance (NQR) spectroscopy is required. Established NQR probeheads with motor-driven tune/match stages are usually bulky and slow, which can be prohibitive if it comes to Bi compounds with low SNR (excessive averaging) and long quadrupolar $T_1$ times. Moreover many experiments yield better results at low temperatures such as 77K (liquid nitrogen, LN) thus requiring easy to use cryo-probeheads. In this paper we present electronically tuned wideband probeheads for bands in the frequency range 20 – 120 MHz which can be immersed in LN and which enable very fast explorative scans over the whole range. To this end we apply an interleaved subspectrum sampling strategy (ISS) which relies on the electronic tuning capability. The superiority of the new concept is demonstrated with an experimental scan of triphenylbismuth from 24 – 116 MHz, both at room temperature and in LN. Especially for the first transition which exhibits extremely long $T_1$ times (64ms) the and low signal the new approach allows an acceleration factor by more than 100 when compared to classical methods.

**Keywords:** nuclear quadrupole resonance, bismuth compounds, wideband cryo-probehead, varactor diodes, tuning and matching, interleaved subspectrum sampling

**Highlights**

- Fast electronically tuneable wideband cryo-coils for NMR and NQR spectroscopy
- electronic tuning/matching allows for very short repetition times
- fast interleaved subspectrum sampling for very short repetition times even at long $T_1$
- strong acceleration of explorative wideband NQR scans by factors of 100 and more

1. Introduction

During the last decades significant efforts have been made to establish magnetic resonance imaging (MRI) in the context of cellular and molecular imaging so as to provide spatially and temporally resolved maps of biomarkers which contain information on pathophysiological processes correlated with cancer and other diseases [1]. So far, however, the relatively low sensitivity requires special signal enhancement strategies, mostly by the administration of an extrinsic tracer or contrast agent (CA) [2] which shorten either predominantly $T_1$ (e.g. gadolinum based paramagnetic chelates) or $T_2$ (e.g. paramagnetic iron oxide nanoparticles) relaxation time of free water protons [3]. In the last years many concepts have been elaborated for "smart" contrast agents that exhibit a change in relaxivity upon activation. Image contrast can be switched in response to physiological alterations such as pH, temperature, metal ions, redox state and enzyme activity [4] [5].
Lately, chemical exchange saturation transfer (CEST) has gained interest as an alternative MRI contrast mechanism. CEST contrast can be switched on and off using a selective radio frequency (RF) saturation pulse but suffers from low sensitivity, specific absorption rate (SAR) restrictions and saturation spill-over effects between bulk protons and protons of the CEST agent due to insufficient frequency separation [6],[7]. Recently new metal-free $T_2$ shortening agents have been reported which rely on nitroxide-based macromolecules [8] and which have the advantage of being free of toxic heavy metals. However, as they represent radicals, their lifetime may become limiting and they are not inherently switchable.

An interesting alternative to the established approaches could be the exploitation of so-called quadrupolar relaxation enhancement (QRE), which means the shortening of $T_1$ by the interaction of protons with quadrupolar nuclei (QN), i. e. nuclei with a spin quantum number ≥1 which possess an electric quadrupolar moment. The effectiveness of QRE for producing MRI contrasts has been proven for the interaction between $^1H$ and $^{14}N$ in the amide groups of proteins at very low magnetic fields [9],[10],[11],[12] but is completely unexplored in the context of extrinsic CAs at clinical fields such as 1.5 or 3T.

QRE, in contrast to most other mechanisms for relaxation enhancement, is frequency selective, because it can only become effective if the proton Larmor frequency comes close to one of the transition frequencies of the QN [12]. This feature opens the possibility to switch on and off QRE by changing the static magnetic flux density $B_0$. As the latter depend on the electronic environment of the QN, QRE is also sensitive to changes of e.g. the chemical bonding structure, which opens many opportunities for chemically selective contrasts.

For several reasons (high spin quantum number 9/2, high natural abundance, rich chemistry, suitable quadrupolar coupling constant $Q_{cc}$) $^{209}$Bi appears as an especially attractive QN. It is considered as comparatively non-toxic and thus compounds are expected to have a higher biocompatibility than e. g. gadolinium based CAs. $^{209}$Bi is being suggested [13] [14] also in x-ray contrast agents which makes it an ideal candidate for dual-mode CAs for CT and MRI with only a single core element. $^{209}$Bi is particularly interesting because there exist several inorganic and organometallic compounds with transition frequencies already close to those required at clinical field strengths, e.g. 3T.

The development of Bi-based contrast agents employing QRE at defined frequencies involves several steps, the detailed discussion of which exceeds the scope of this paper. In any case, however, the transition frequencies of the synthetized compounds must be determined, which, in most cases, can be done easiest with zero-field nuclear quadrupole resonance spectroscopy (NQRS). NQRS has several applications in solid state physics e.g. for the investigation of chemical bonds and crystal structures [15][16], the real-time detection of specific toxic species (e.g. arsenic) in minerals [17] or the detection of explosives [18] and drugs [19] and the characterisation of pharmaceutical products [20]. Most current applications refer to $^{14}$N with very low relevant frequencies up to only several MHz. In contrast, the characterization of suitable Bi compounds requires NQRS in the VHF range, because the $^{1}$H Larmor frequencies of most clinical MRI systems range from several tens of MHz (e.g. 64 MHz@1.5T) up to 130MHz (3T) or even 300 MHz (7T). The identification of still unknown peaks in newly synthetized $^{209}$Bi compounds and their full spectral characterization requires the scanning over many tens of MHz due to the presence of at least four transition frequencies in spin-9/2 nuclei. Thus the typical resonators in NMR probeheads cannot be used because of their small bandwidth. In order to circumvent this problem, commercially available probeheads are equipped with stepper motor drives which mechanically control the tuning and matching capacitors. Though robust and low-loss such implementations have their disadvantages, i. e. they are comparatively large, heavy and slow. As will be shown, in the case of samples with low SNR and long $T_1$ times such an approach renders NQRS extremely tedious due to the need for excessive averaging with long repetition times. This problem can become stringent when measuring at low temperatures, e.g. in liquid nitrogen (77K), which is sometimes necessary in order to increase the SNR by population enhancement and, in favourable cases, a decrease of the transversal relaxation rate.

The aim of this paper is to present a very efficient solution which allows for a fast wideband scanning strategy based on a previously developed hybrid probehead which combines wideband transmit (TX) with electronically tuned and matched narrowband receive (RX) [21]. First we describe the modification of the hardware for operation at temperatures down to 77K (liquid nitrogen) and second we introduce an interleaved subspectrum sampling (ISS) sequence which relies on the rapid electronic tuning capability. Finally experiments with two promising $^{209}$Bi-containing core-compounds for QRE will show the usefulness of the presented methodology.

## 2. Methods

*Fast scanning strategy*

In NMR spectroscopy bands of several 100 kHz can be excited by an appropriately controlled RF pulse sequence like e.g. the double-frequency sweep method [22] or related concepts [23] [24] and the WURST-QCPMG sequence [25]. However, as described in the excellent review about wideline acquisition by R. Schurko in [26], bandwidths of more than a few 100kHz require a frequency stepping method, which is especially true for explorative NQRS scans over tens of MHz. One of the limiting factors for the overall duration of an experiment is the repetition time $T_R$, which in the case of FID or spin echo sequences is typically greater than the longitudinal relaxation time $T_1$. To increase the SNR mostly N spectra have to be averaged, and therefore the overall acquisition time per frequency is $N*T_R$. Samples with long $T_1$ thus require long acquisition times.

Frequency stepping basically consists in repeating the above procedure for M different center frequencies into which the whole frequency span FS is subdivided [27], [28]. The resolution $\Delta f=FS/M$ is a function of the sampling rate, the acquisition window length and the minimum expected peak width. Basically there are 2 different possibilities for the frequency stepping strategy: (1) At every pulse frequency N datasets are acquired and averaged before hopping to the next frequency (2) Complete wideband spectra are acquired by sweeping N times through the M frequencies and doing the averaging afterwards. Strategy (1) is common with commercial NMR consoles while strategy (2) is the way many network analyzers (NWA) do their job.

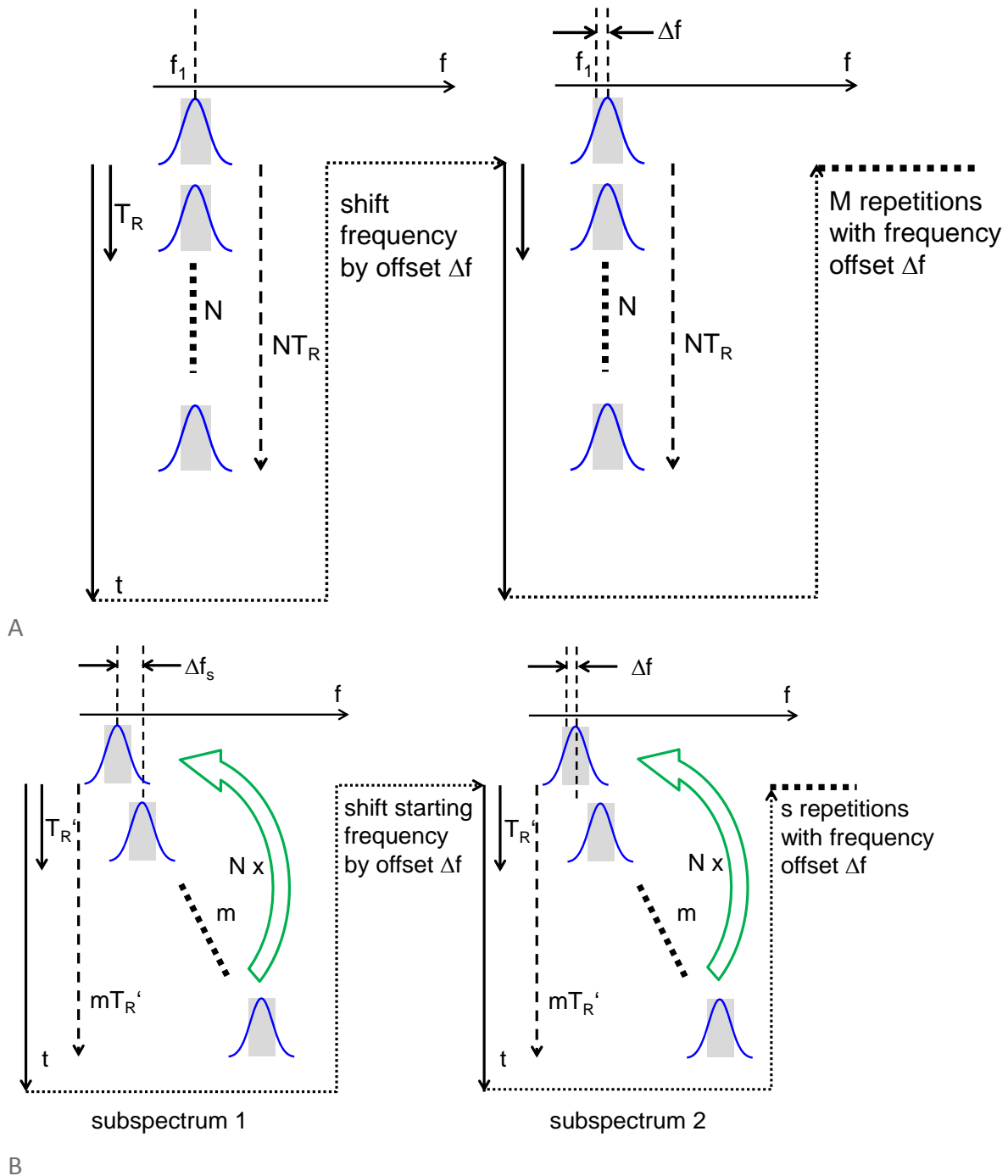

*Fig. 1. A: Classical sampling strategy: N data points are averaged per loop, M loops with frequency offset $\Delta f$ are accomplished. $T_R \geq T_1$*
*B: Fast interleaved subspectrum sampling: s coarsly spaced, interleaved subspectra are subsequently acquired in s=M/m loops with frequency offset $\Delta f$. Within each loop averaging over N repetitions is accomplished (symbolized by the green curved arrows). $T_R' \geq T_1/m$*

At a first glance the two strategies may appear equivalent. However, with strategy (2) the whole procedure can be accelerated significantly by interleaved subspectrum sampling (ISS), as illustrated in fig.1. Panel A shows the classical strategy, where N data points are acquired per acquisition loop and a simple spin echo sequence is assumed. Then the frequency is shifted by $\Delta f$ and the procedure is repeated M times, giving a total measurement time of $T_R*M*N$, where $T_R \geq T_1$.

Fig.1B illustrated the ISS, where FS is subdivided into m≤M frequency points the separation $\Delta f_s$ of which corresponds at least to the bandwidth $BW_{TX}$ of the TX pulse. In this case we have $\Delta f_s = FS/BW_{TX}$ and consequently $m=M*\Delta f/\Delta f_s$. In one acquisition loop N sub-spectra are acquired by sweeping through the m frequency points before averaging them. As each TX pulse only excites spin populations within its bandwidth, populations at the neighboring TX frequencies cannot be saturated and thus the available recovery time for each population is $m*T_R'$ instead of $T_R$. The requirement $T_R \geq T_1$ can be relaxed to $T_R' \geq T_1/m$, thus allowing for the much shorter repetition times $T_R' = T_R/m$.

In total a number s=M/m of such loops must be accomplished in order to re-compose the original spectrum from the coarsely spaced sub-spectra whereby in each loop i the respective starting frequency must be shifted with respect to the previous loop i-1 by the wanted resolution $\Delta f$ (see fig. *1*B, subspectrum 2). In this way the total number of frequency points acquired is again M with a resolution of $\Delta f$.

However, the total measurement time $T_{meas}$ after assembling and averaging of all subspectra is then $T_{meas} = T_R'*m*N*s = T_R'*M*N$ which is by a factor of m less than for the classical case. Because the number of required datasets is the same for both methods, the maximum acceleration factor is m.

For a better illustration assume a sample with $T_1$ = 10 ms with a weak signal which requires N=1000 averages. The explorative scan should cover FS=12 MHz and the resolution is set to $\Delta f$= 20 kHz, thus yielding M=600 frequency points. A spin echo sequence with rectangular pulses $T_P(90°)$=10μs, $T_E$=100μs, $T_{DAQ}$= 50μs is applied. The bandwidth of such a pulse for more than 50% effectivity (FWHM) in the small flip angle limit is $1.21/T_P$ [29] i. e. roughly 120kHz. For simplicity let $T_R$ be equal to $T_1$, which yields $T_{meas}$=1000 x 600 x 10 ms =6000s = 100 min for the classical scan.

Acquisition with ISS requires m = 12MHz/120kHz = 100 frequency points per subspectrum and s=6 subspectra for splitting $\Delta f_s$ into 20 kHz steps. Thus $T_R'$ can be chosen as short as $T_1/m$=100μs, which reduces $T_{meas}$ to 1 min, i. e. one hundredths of the time needed for the classical technique.

The classical technique works fairly well with the slow tuning/matching elements driven by stepper motors, because the averaging of N datasets is accomplished before every re-tune/re-match step. If the averaging lasts much longer than the re-tuning procedure, there is not much delay. ISS, in contrast, requires re-tuning after every $T_R'$ and is thus only possible with fast electronically tuned probeheads. Thus the tuning/matching elements must settle within $T_R'$, i. e. in less than 100μs for the shown example. This is clearly impossible with motor drives but very well feasible with varactors. There are, however, two limitations for the acceleration factor: (1) For a spin echo sequence $T_R'$ cannot be shorter than $T_P/2+T_E+T_{DAQ}/2$, whereby $T_E$ is the echo time and $T_{DAQ}$ the window length for data acquisition. (2) Shorter $T_R'$ increases the duty cycle of the RF pulses and thus the average power dissipation in the probehead. Therefore in practice the shortest $T_R'$ used in our probeheads was 500µs.

*Probehead circuitry*

The probehead electronics is an advanced version of the one in [21] where the basic concept has been described in detail. Thus only the essential ideas shall recapitulated very briefly. As shown in fig. 2 the circuit is subdivided into the passive TX network and the electronically tuned and matched RX circuitry.

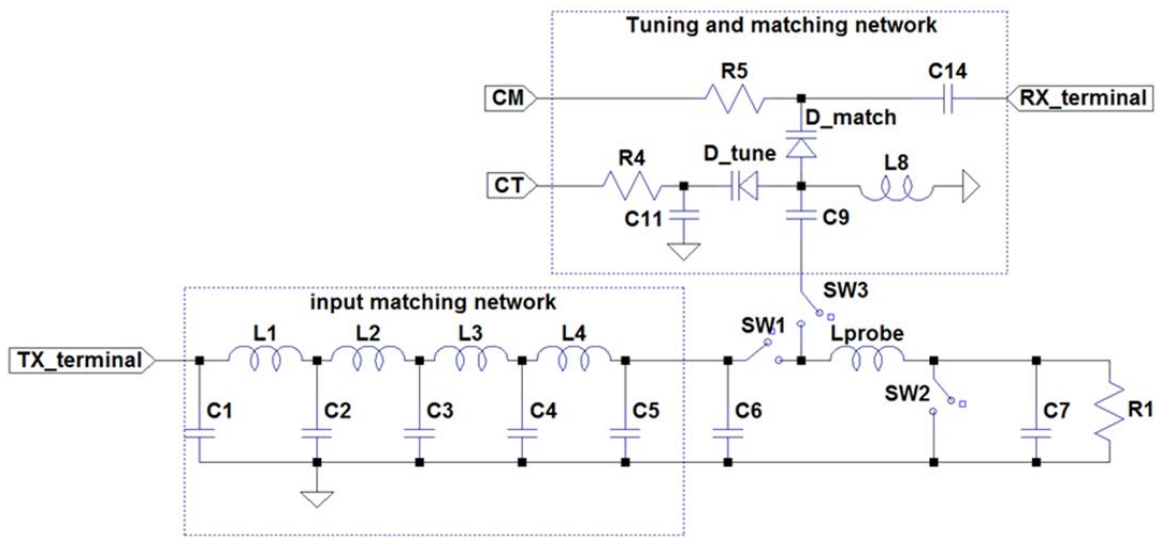

*Fig. 2. Principle of the TX/RX matching circuitry (from [21]). During TX switch $SW_1$ is closed while $SW_2$ and $SW_3$ are open. During RX all switch states are reversed. The numbering of the components is not strictly monotonic because some parts have been left out for the sake of simplicity. It is, however, consistent with the detailed schematic in fig. 4.*

During TX switch $SW_1$ is closed while $SW_2$ and $SW_3$ are open and hence power is passed to the termination resistor $R_1$ via a reactive network which represents a lumped exponential transmission line (LETL). It consists of the input matching network and the pi-section $C_6, C_7$ and $L_{probe}$ and transforms the termination resistance $R_1$ into the 50Ω input impedance over a large bandwidth, thus providing maximum current in the probe coil $L_{probe}$. During this phase the tuning/matching network is isolated from the circuitry by the open $SW_3$. During RX $L_{probe}$ is isolated from the power path by opening $SW_1$ and connected to the tuning/matching section by closing $SW_2$ and $SW_3$. The NMR signal can now be received in resonator mode with high quality factor Q and passed to the 50Ω port 'RX_terminal'. Tuning and matching is achieved with the varactors D_tune and D_match which are controlled via the reverse bias voltages provided at the ports CT and CM and which can settle fast enough for ISS.

For cryo-applications we have modified the previous probeheads according to the cross sectional drawing in fig 3. It consists roughly of a printed circuit board (PCB) with the RF electronics and connectors as well as the probe coil assembly. The latter is supported by a hollow PTFE tube which carries at its tip the probe coil and along its outer surface a stripline which connects the coil with the PCB. A thin outer plastic tube protects the assembly and serves as a guide for cooling gas as well as a certain thermal insulation between the grounding copper of the stripline and the coolant. The probe coil can be either immersed directly into liquid nitrogen (LN) or inserted into a cryostat. The cold vapor of the LN or any other cooling gas can escape through the inner bore of the PTFE tube and removed at the nozzle protruding from the top side of the PCB. A thermal insulator made from polystyrene foam protects the PCB from excessive cooling and both the bottom of the PCB as well as the nozzle carry electric heaters so as to prevent the condensation of water along cold parts which are exposed to air humidity.

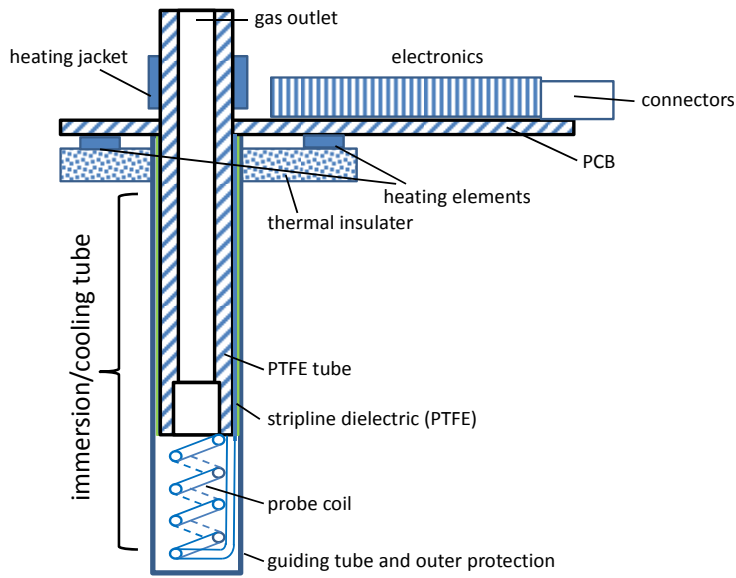

Fig. 3. Lateral cross-section of the probehead,

The choice of the values of all the inductances and capacitances of the LETL must obey a strict rule which follows from defining the center frequency $f_c$, the transform ratio $R_1/50\Omega$ and the number of pi-sections. The TX bandwidth $BW_{TX}$ is then typically between 40 and 50% of $f_c$, depending on the definition of the useful bandwidth. In our case we define $BW_{TX}$ as the one within which the magnitude of the input reflexion factor $S_{11}$ at the TX terminal does not exceed -10dB, a value which is tolerated by most RF power amplifiers. The choice of $R_1$ determines the Q-factor of the probe coil, because, as detailed in [21], the inductance of the latter is proportional to $R_1$. With a number of pi-sections ≥ 4 and $R_1$ ranging between 10 Ω and 200Ω the relationship between $L_{probe}$ and $R_1$ is approximately

$$L_{probe} \approx \frac{R_1}{2\pi f_c} \qquad (1)$$

and

$$Q \approx \frac{f}{f_c}\frac{R_1}{r} \qquad (2)$$

where f is the actual operating frequency and r is the loss resistance of the probe coil. In contrast to [21] $R_1$ was set to 100Ω, which implies comparatively high $L_{probe}$ and Q as well as an up-transformation of the voltage by a factor of $\sqrt{\frac{R_1}{50\Omega}}$ i.e. 1.41. The current, in turn is transformed down by the same factor. Both effects have advantages and disadvantages. On one hand some elements, especially the capacitors $C_6$ and $C_7$ and the switches $SW_2$ and $SW_3$ must withstand comparatively high voltages (approx. 141 $V_{peak}$ at an RF power of 100W), but on the other hand the resistive losses and the thermal stress of $SW_1$ decreases by a factor of 2.

The detailed schematic of the RF circuitry is shown in fig. 4. For comparison, all common components are numerated in the same way as in fig. 2. Component values have been omitted intentionally because they depend on the individual frequency bands. Instead design rules are being discussed.

As in [21] PIN diodes (MA4P7104F-1072T, MACOM) were used for $SW_2$ and $SW_3$ but $SW_1$ was replaced by the crossed diodes $D_1 - D_4$ with low reverse recovery time (BAS316, NXP). A quadruple was used instead of the pairs commonly described in textbooks in order to increase the allowed power dissipation. Crossed diodes are very common (cf. e.g. [30]) but their downside is the higher parasitic capacitance (here typ. 2.5 pF per quadruple) when compared to a reverse-biased PIN diode of similar power ratings (here 0.7 pF). Coupling between the TX matching network and the probe coil via these parasitic capacitances can significantly lower the Q-factor during RX, especially at high frequencies (70MHz and above). Thus for the highest frequency band each diode was replaced by a series of two or three diodes so as to reduce the total capacitance to less than 1 pF. The crossed switching diodes $D_8$ and $D_7$ protect the sensitive varactors and the preamplifier following the RX port from high residual RF signals during TX.

$C_8 - C_{14}$ serve as RF shortcuts in the respective frequency bands. Their values should be as high as possible, but all of them are limited by timing considerations. $C_{11}$, $C_{12}$, $C_{13}$ together with resistors $R_4$ and $R_5$ provide the reverse biasing of the varactors and must be chosen such that the bias voltage settles to within 1% in less than the minimum wanted repetition time $T_R'$, e.g. a few 100 µs for ISS. $R_5$ must be significantly higher than the typical RX load impedance of 50 Ω, e. g. 1 kΩ while $R_4$ must be significantly higher than the reactance of $C_{10}$, typically several kΩ. Resistors were used instead of the usual inductive chokes because they do not produce spurious ringing after step changes of the bias voltages.

In the spectrometer the voltage at 'HV_PIN' is 250 V during all TX pulses (reverse bias) and -5V (forward bias) during the rest of the time. The resistors $R_2$ and $R_3$ in the biasing networks $R_2/L_6$ and $R_3/C_{10}/L_7$ limit the PIN diode forward bias currents to ca. 100mA and thus have 39Ω, each. $L_6$ and $L_7$ serve as RF chokes while $L_8$ provides a common GND reference for $D_6$ and both varactors. The inductances must be chosen significantly higher than that of $L_5$ in order not to change the properties of the tank circuit. Moreover $L_7$ and $L_8$ must have high Q factors, because they lie in parallel to the probe coil from the RF viewpoint and their losses contribute directly to the resonator losses. These coils were thus fabricated with air core and comparatively thick (0.4 mm) enamelled copper wire. In this context $C_{10}$ plays an important role as RF shortcut so as to decouple the ohmic loss $R_3$ from $L_7$.

$C_8$ and $C_9$ decouple the PIN diode bias paths from the RF path. They must be rated for ≥ 250 V and their capacitances must be kept as low as possible, providing still good RF passage but limiting the switching currents during the edges of the 250 V pulses to values which do not significantly afflict the rise time and the power demands for the control pulse source.

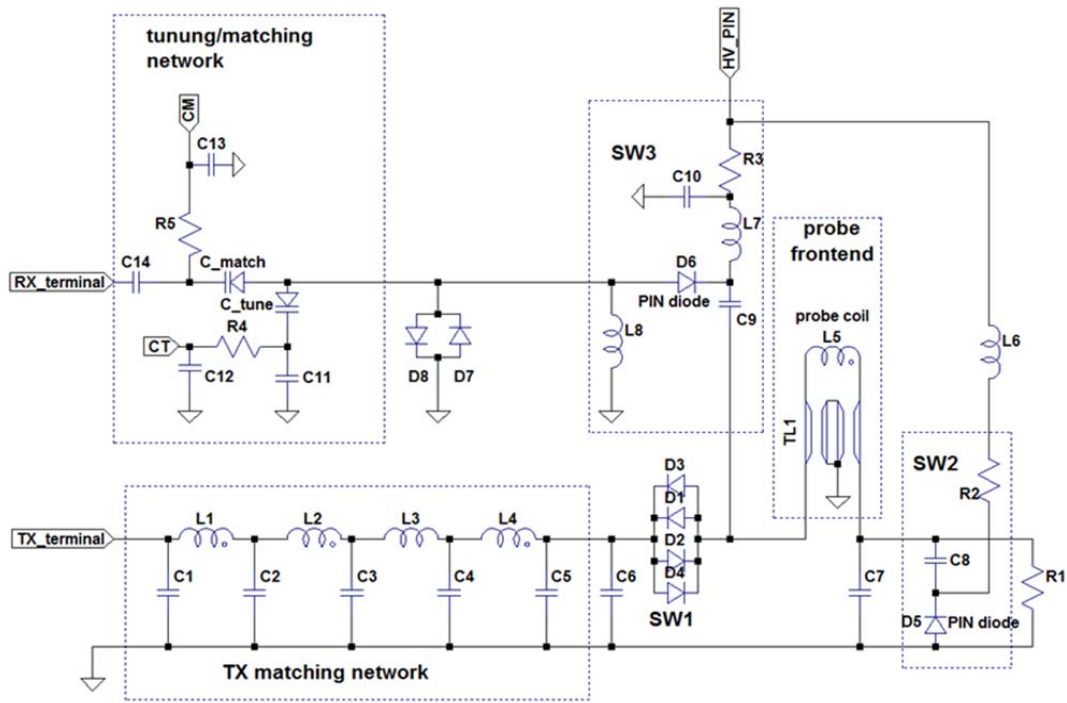

Fig. 4. Detailed schematic of the RF section of the PCB.

The design of the connection line $TL_1$ between the probe coil and the PCB deserves special attention because it requires a trade-off between partially conflicting goals:

*(A) Thermal viewpoint:*

(i) During cryo-applications a longer PTFE tube allows for deeper insertion into LN and/or for better thermal insulation between the PCB and the LN.

(ii) The connection line should have a small cross-section in order not to convey too much heat flux from the tip to the PCB, which increases consumption of coolant and the danger of condensation on the PCB

*(B) RF viewpoint:*

(iii) The connection line along the tube adds parasitic inductance in series with the useful part $L_5$ of the probe inductance, i.e. the one which actually receives the signal. As the total inductance is fixed according to eq 1 for a certain frequency band, the parasitic amount reduces the RX sensitivity and should be kept minimal. The parasitic inductance decreases with the area enclosed by the line conductors and typically increases with tube length.

(iv) The Q-factor is inversely proportional to the ohmic losses of $L_{probe}$. Therefore the cross-sectional circumference of the conductors of both $L_5$ and $TL_1$ should be large in order to account for the skin effect. Typically this also increases the cross-section of the respective conductors.

Requirements (i) and (iii) as well as (ii) and (iv) are contradictive, respectively, and thus a compromise must be found. A good solution is an edge-coupled symmetric microstrip line, because it allows for large surface/cross-section ratio of the conductors thus serving both (ii) and (iv). The gap between the strips should be small in order to satisfy (iii). The dielectric should be thin enough in order to prevent the current to concentrate close to the strip edges due to the proximity effect which would increase the losses. On the other hand the thickness must exceed a certain limit to prevent excessive capacitance per length which adversely affects the transformation properties of the line. In the simplest case (electrically short line) the even capacitances between strips and ground add to $C_6$ and $C_7$, respectively, forcing their values to be reduced. The maximum allowable line length is reached when either of them is replaced entirely by the line capacitance which can be calculated from $Z_0$ and the phase velocity c:

$$Z_0 = \sqrt{\frac{L'}{C'}}; \quad c = \frac{1}{\sqrt{L'C'}} \tag{3}$$

$$C' = \frac{1}{cZ_0}, L' = \frac{Z_0}{c} \tag{4}$$

C' and L' denote the per-length capacitance and inductance, respectively. Therefore the maximum length can be approximated as

$$l_{max} = \frac{C_7}{C'} \tag{5}$$

Parametric simulations with LTSpice (Linear Technologies) were carried out to find out the best $Z_0$ at a certain desired length of the line so as to achieve maximum bandwidth both at TX and RX. Then the stripline geometry was calculated with a web-based stripline calculator (https://www.eeweb.com/toolbox/edge-coupled-microstrip-impedance) for the required $Z_0$. The thickness of the dielectric was chosen with 0.5 mm for practical reasons. A PTFE sheet was chosen for the dielectric due to its low losses and low capacitance per length. As conductors we used self-adhesive copper foil with a thickness of 50 µm and a strip width of 3 mm. In order to minimize thermal contact with the coolant, the outermost copper part (in our case the GND layer) was insulated by a plastic tube which at the same time serves as a guide for the cooling gas around the probe coil (see fig. 3). As a general rule increasing the frequency leads to shorter tubes because the capacitors must decrease inversely proportional to $f_0$.

Let us give a practical example: Setting $f_c$ to 95 MHz in order to get a useful TX band between 70 and 120 MHz requires a $C_7$ of ca. 17 pF. A reasonable value for $Z_0$ is 30Ω and c in a PTFE dielectric is approximately $2*10^8$ m/s. Thus we obtain a L' of 150 nH/m and a C' of 167 pF/m. The maximum theoretical length of the tube is thus 9.7 cm. However, due to additional parasitic capacitances on the PCB, this value must be reduced so that in practice the typical length is 6-7 cm only. The effective parasitic inductance is then 17 nH which is less than 10% of $L_5$ and thus practically negligible. Going down in frequency to an $f_c$ of say 30 MHz allows for a maximum tube length of >30 cm which is more than sufficient for most applications.

At high $f_c$ a somewhat longer tube is possible at the cost of a reduced bandwidth with a modification of the circuit according to fig. 5. The difference to fig. 4 is the replacement of $SW_3$ by a passive RX protecting network consisting of two series connected lumped λ/4 line equivalents and two pairs of clamping diodes $D_7/D_8$ and $D_{11}/D_{12}$. This circuit is very common in NMR coils, see e.g. [30],[31]. The first λ/4 equivalent is formed by $L_7$, $C_9$ and one half of $C_{10}$ while the second one, less obviously, is built up by $L_9$, the second half of $C_{10}$ and a part of the tuning capacitance $C_{tune}$. The resonance frequency of this network is placed approximately in the center of the desired scanning bandwidth. Exactly at resonance the function of this circuit is obvious: During TX $D_7/D_8$ form the usual short-circuit which is transformed to an open end at the connection point with the probing coil and thus does not perturb the TX pulse. In turn the RX section is protected by the clamping diodes. The second pi-section serves both for further reduction of the residual voltage during RX and, more importantly, reverts the action of the first λ/4 element by extending it to a λ/2 line. Therefore, during RX, the tuning varactor 'sees' directly the probe coil. As simulations show, this mechanism is also effective off-resonance within a certain bandwidth. After tuning $C_9$, $C_{10}$ and the transmission line parameters of $TL_1$ in the simulation good tuning and matching can be achieved over a bandwidth of approximately 20% of $f_0$. During TX an $S_{11}$ of at least -12 dB can be achieved within this band. The reason for the larger allowable tube length is the fact that in this configuration $C_9$ is in parallel to $C_6$. Thus the transmission line capacitance is allowed to 'consume' also part of $C_9$ and still provides a fair $S_{11}$ during TX. In our practical implementation we could reach a tube length of 10 cm, which was sufficient for all our applications. As the bandwidth is considerably reduced with respect to the original implementation, we provided two different RX protecting networks with different resonance frequencies onto the same PCB and made them selectable by two jumpers. This way the band between 79 and 116 MHz can be covered with a single probehead, which corresponds to approximately 40% of the central TX frequency. One major disadvantage of this design are the additional resistive losses in the coils $L_7$ and $L_9$ which reduce the SNR by about 20% compared to the original circuit. In order to keep the losses as low as possible the coils were made from comparatively thick (0.8 mm diameter), silver-plated copper wire.

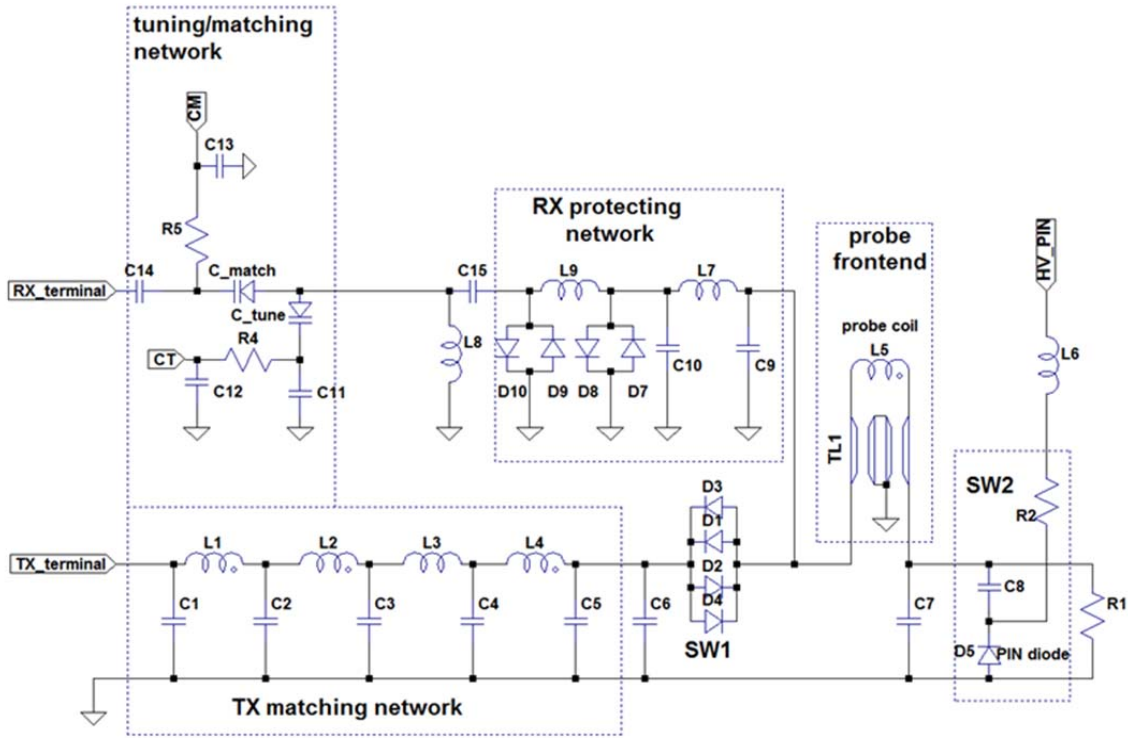

*Fig. 5. Detailed schematic of the alternative circuit with passive RX protecting network.*

*Spectrometer*

In principle the probehead can be used in combination with commercial spectrometers. In one implementation we used it with a Tecmag Scout (Tecmag, Inc., USA) using the classical technique. However, as it is difficult to program this device for the ISS, we apply this configuration only for accurate relaxometric measurements where the advanced pulse programmer of the commercial console is required. For fast explorative scans we employ the custom-made spectrometer described in [21] which relies on the network analyzer (NWA) ZWA03 (Rohde&Schwarz). For details of the data acquisition, the calibration of the tuning/matching voltages and the console program written under LABVIEW please refer to [21].

*Coil characterization*

The matching of the TX coil was measured by scanning the $S_{11}$ parameter at the TX port while switching the HV_PIN control line to TX-mode and bridging $SW_1$ by a jumper. Measurements were done with the same NWA which was used in the spectrometer. As the useful TX bandwidth we defined the frequency range within which the return loss remained below -12dB. Moreover the Q factor was measured by measuring the $S_{12}$ parameter between two magnetically decoupled coils (double-loop coil) which were weakly coupled to the probe coil so that a clear resonance peak could be obtained in RX mode. Q was calculated from the 3-dB bandwidth of the peak at 10 different resonace frequencies $f_r$ within the entire RX bandwidth. The settling times of the control voltages for the varactors and PIN diodes were checked with an oscilloscope (Tektronix TDS3014B) while measuring the rise-times at the respective points. All settling times were found to be within the requested limits.

In total we built three different probeheads according to fig. 4 and one prototype of the alternative configuration in fig. 5. Their frequency ranges and Q-factors in the middle of the respective band ar summarized in table 1.

| Name | Frequency range (MHz) | Q ($f_{middle}$) | Tube length (mm) |
|---|---|---|---|
| A3 | 63-116 | 50 | 62 |
| B3 | 34-65 | 51 | 105 |
| C3 | 24-41 | 52 | 137 |
| A4 | 79-116* | 70 | 90 |

Table 1: Frequency ranges and Q-factors ath the central frequency $f_{middle}$ of the respective band. *band was split into two subbands from 70 – 95 and from 95 – 116 MHz.

*Validation experiments at room temperature and in liquid nitrogen*

In the framework of the research project which motivates this paper (see acknowledgement) we investigated a family of organometallic $^{209}$Bi compounds which play a central role as starting materials for a potential QRE based contrast agent. From this family we chose triphenylbismuth (Sigma Aldrich) for the evaluation of our system. For comparison with our previous probeheads [21] we also measured spectra of ZnBr$_2$ which contains Bromine (Br) as QN with spin quantum number I=3/2. The solid powders were filled up to a height of 2 cm into glass cuvettes with an inner diameter of 6.5 mm and pressed gently so as to maximize the filling factor. Then the cuvettes were closed with a plastic stopper, placed inside of the coils and secured. The temperature of the samples was measured with a commercial thermoelement type K coupled to a multimeter TE.ELECTRONIC MS8264. In order to keep the temperature widely constant a continuous thermostatised airstream was passed through the PTFE tube of the probehead from top. Thermostatisation was achieved with a Peltier Temperature Controller (NMR Service GmbH, Germany). After settling to the target temperature (27.5°C for triphenylbismuth and 25°C for ZnBr$_2$) the ISS was used to scan over the whole accessible range between 20 and 116MHz in sub-bands with FS = 15MHz or 30 MHz and m=150 or 300, respectively, depending on the individual coil bandwidths. In all cases s=10 subspectra were acquired, corresponding to a $\Delta f_s$ of 10 kHz. N was set either to 100, 200 or 500, depending on the peak amplitude within the respective sub-bands. A spin echo sequence was used with $T_E$=80ms and $T_R$'=1.5ms. The pulse duration was chosen with 14µs for coil A3 and 8µs for coils B3 and C3 while adapting the pulse power so as to have approximate flip angles of 90°.

The experiments were repeated after plunging the respective probehead into a Dewar with LN and allowing the sample to fully equilibrate to -196.15°C (77K) for 1min. A similar experiment was carried out for a sample of $ZnBr_2$ in order to show a more complex, but narrower spectrum for comparison.

3. Results

Fig. 6 shows the complete wideband spectra obtained from triphenylbismuth at 27.5 and -196.15°C in a single plot. The data from the three coils A3 – C3 were concatenated into 'panoramic' full datasets and plotted as raw data. The SNR of identified peaks was calculated as the ratio between the peak amplitude and the RMS value of the noise in the baseline interval. Peak amplitudes are given in $\mu V_{RMS}$ received at the input of the low noise preamplifier. One can clearly distinguish the four narrow peaks labelled TR1 – TR4 which correspond to the transitions $|3/2\rangle \rightarrow |1/2\rangle$, $|5/2\rangle \rightarrow |3/2\rangle$, $|7/2\rangle \rightarrow |5/2\rangle$ and $|9/2\rangle \rightarrow |7/2\rangle$, respectively, at the frequencies 30.64; 56.44; 85.45 and 114.0 MHz at -196.15°C and 29.76, 55.21, 83.51 and 111.4 MHz, respectively, at 27.5°C. The most prominent features are the strong signal enhancement (and thus SNR gain) by a factor of up to 4 and the upward shift of the transition frequencies when decreasing the temperature from 27.5°C to -196.15°C. All transition frequencies at room temperature are in excellent agreement with those published in the Landoldt-Börnstein database [32].

Though the figure enables a very accurate localization of the transition frequencies, the amplitudes of the peaks must be interpreted with caution for the following reasons:
(1) Different coils have slightly different geometries and winding densities, therefore their filling factor and magnetic field distribution is somewhat different, leading to different sensitivity.
(2) The magnetic flux density varies by up to 20% within the coil bandwidth and thus the flip angle is not optimal for all frequencies when scanning over very large bands such as 30 MHz. (The current spectrometer implementation does not allow for an automatic compensation of this effect).
(3) The $T_2$ relaxation times vary considerably over the transitions TR1 – TR4 as well as with temperature, ranging from as short as 80 $\mu s$ for TR1@27.5°C to 822$\mu s$ for TR4 @-196.15°C. Therefore the $T_2$ weighting is different for all peaks. The quantification and discussion of the relaxation rates, however, is beyond the scope of this paper and shall thus be left to a separate publication.
(4) Sample positioning inside of the coil may have been subject to a certain variability.

Despite all these imperfections the relative amplitudes correspond approximately to those which can be expected from theoretical calculations of the quantum-mechanical transition probabilities, the Boltzmann populations, the transverse relaxation time constants $T_2$ and the induction law. The variations of the noise floor between different coils is due to the different levels of averaging (100, 200 or 500).

The total scanning time was approximately 45 min per complete spectrum, i. e. 90 min for fig. 6. The SNR varied between 51 (TR1@27.5°C) and 578 (TR3@-196.15°C).

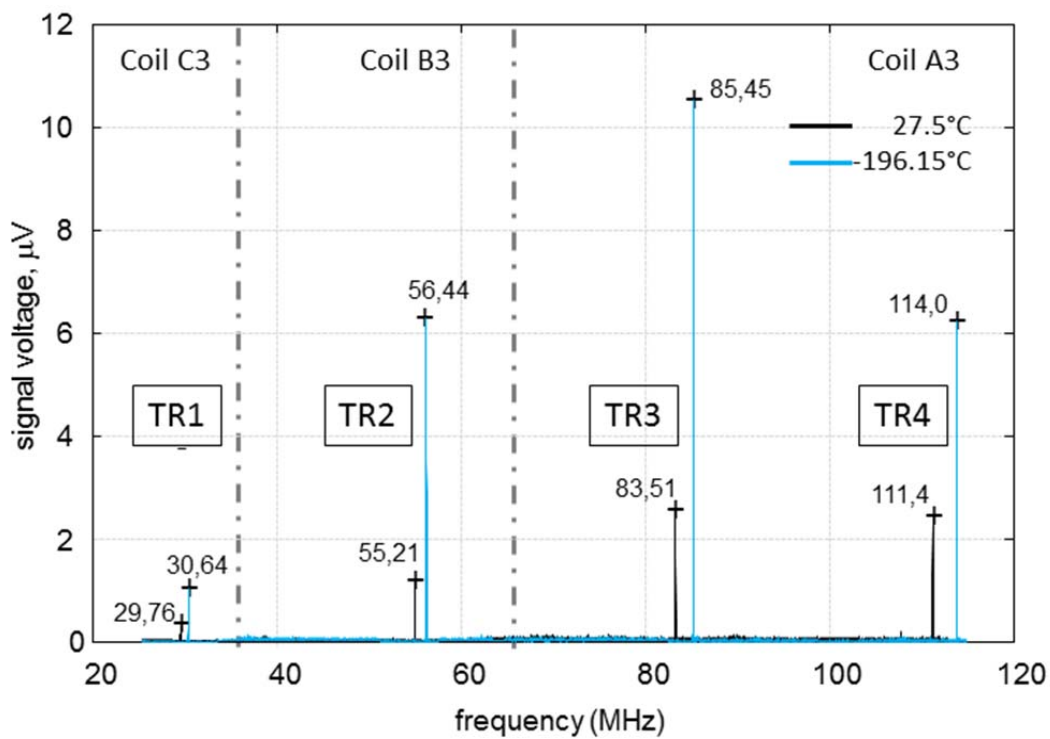

*Fig. 6. Complete wideband spectra of triphenylbismuth at -196.15°C (blue) and 27.5°C (black), showing all four transitions TR1-TR4, respectively. The numbers beneath the peaks denote the respective transition frequencies in MHz. The dash-dotted vertical lines separate the bands of the three coils.*

Fig. 7 shows two superimposed spectra of anhydrous $ZnBr_2$ at two different temperatures, i.e. 25°C and -196.15°C. As Br has a spin quantum number of 3/2 the quantum-mechanical rules allow only for one single transition. However, both spectra contain two triple peaks labelled as A1-A3, B1-B3 and A1'-A3', B1'-B3', respectively, whereby primed labels refer to the measurements at room temperature. The reason for the multiple peaks is twofold: First there exist two different Br isotopes namely $^{79}Br$ (label B, triple peaks above 95MHz) and $^{81}Br$ (label A, triple peaks below 85MHz). Second: Each triplett consists of the contributions from three different crystallographic sites in the $ZnBr_2$ crystallites, whereby the central peak is approximately twice as high as the two satellite peaks, see [33] for more details. The measured peak frequencies are given in table 2:

| Temperature | -196.15°C | | | 25°C | | |
|---|---|---|---|---|---|---|
| site | 1 | 2 | 3 | 1' | 2' | 3' |
| Isotope A | 81,43 | 83,08 | 84,14 | 79,75 | 81,40 | 82,33 |
| Isotope B | 97,47 | 99,48 | 100,7 | 95,46 | 97,44 | 98,55 |

*Table 2: Transition frequencies in MHz for the different Br isotopes and sites in $ZnBr_2$.*

The data are in agreement with those published in [34].

Again the strong upshift in frequency as well as the decrease of the amplitude with decreasing temperature is clearly visible.

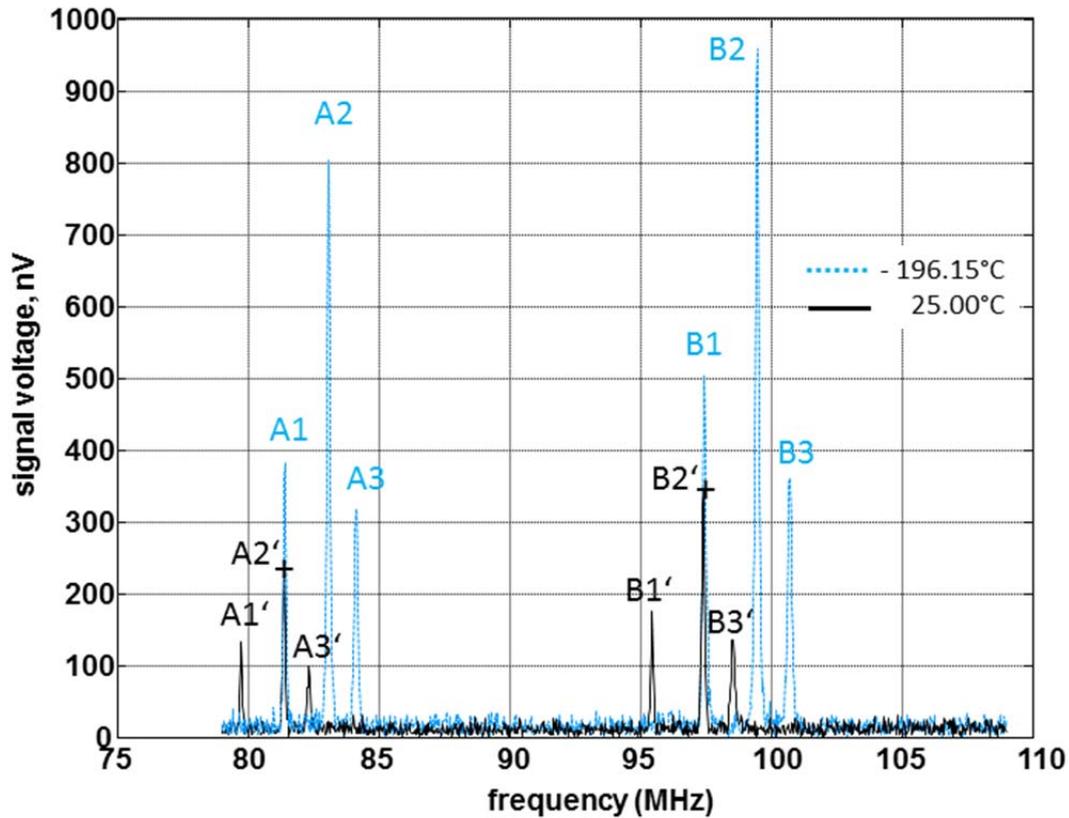

Fig. 7. Wideband spectrum of $ZnBr_2$ Anhydrid, 1000 points, 10 averages @-196.15°C (blue/dashed) and 25 averages at 25°C (black/solid), IF BW 10kHz, span 30 MHz, classical scan, spin echo with Tp = 22 µs, $T_E$ =200 µs, $T_R$=3000 µs.

**Discussion:**

We have presented a fast spectrometer for nuclear quadrupole spectroscopy in the VHF range which relies on special electronically tuneable wideband cryo-probeheads and a rapid scanning strategy (interleaved subspectrum scanning, ISS).

Complete wideband spectra of triphenylbismuth and $ZnBr_2$ have been shown. The advantage of the ISS protocol was demonstrated for the identification of transition1 in triphenylbismuth at -196.15°C, the $T_1$ time of which is 64ms and the signal of which is comparatively low. With the classical scan technique $T_R$ typically would have to be set to 3 $T_1$

, i. e. nearly 200 ms in order to preserve maximum SNR. With the same frequency spacing and averaging as in our experiments the scan over the band of coil C3 would require approximately 41 hours, while the ISS took only 18 min at $T_R'$ of 1.5 ms, providing an acceleration factor of 136. Thus the classical technique may become prohibitively tedious for long $T_1$ at low SNR while ISS together with fast electronic tuning is very well feasible under these conditions.

There is of course also a limitation of ISS: The maximum allowable power dissipation limits s and thus the shortest allowable $T_R'$. In our experiments and with our coils the safe limit was given with a $T_R'$ of around 1.5 ms. If $T_1$ is shorter, the ISS does not offer any significant advantage. In the case of triphenylbismuth at room temperature $T_1$ is only 2915 µs for transition 1 and thus with our $T_R'$ of 1.5 ms the acceleration factor reduces to 2-3 and at transition 4 ($T_1$ = 833ms) there is practically no speedup any more.

The comparison so far is not completely fair because the Q-factor of a classical probehead with a simple LC-resonator can be considerably higher than the one reached with our design (typically 40-50). The main reason are the varactor diodes with their low and frequency-dependent loss factors. The Q-factor could still be increased at the cost of bandwidth, but in general the LC tank is superior, usually up to a factor of 4. Therefore, theoretically, the acceleration factor is reduced by sqrt(Q).

However, with the classical resonator there arises another severe limitation when it comes to samples with very short $T_2$: Ringdown artefacts after TX requires a certain blanking time between the end of TX and the start of data acquisition, which prevents the detection of peaks with very fast $T_2$ relaxation. Fortunately, unlike in NMR, acoustic ringing [35] is not an issue because of the absence of a static magnetic field but all the other types of synchronous transient perturbations affect the data. In an ordinary resonator roughly 23 time constants $\tau$ are necessary to come down from 100 $V_{pp}$ during TX to less than 10nV (see also [36] [37]). As derived in [38]

$$\tau = \frac{2Q_{loaded}}{\omega_0} \qquad (6)$$

whereby $Q_{loaded}$ is half the unloaded Q. Assuming an unloaded Q of 200 and a resonance frequency of 90 MHz (middle of the band of coil A3) we thus get t = 350 ns and a necessary blanking time of 8 µs. In contrast, the Q-factor of our probehead is switched from approximately 2 during TX to 50 during RX thus yielding two different time constants $\tau_{TX}$ and $\tau_{RX}$ in subsequent phases. Eq. 6 then yields a $\tau_{TX}$ of ca. 7ns in coil A3 and 19.6ns for coil C3 while $\tau_{RX}$ is 75ns and 500ns, respectively. After the end of the RF pulse we implemented a short safety interval $T_W$ of 1µs before switching to RX mode, thus leaving enough time (more than 50$\tau_1$ for coil C3) to damp any residual oscillation below detectability when switching to RX. This mechanism can be considered as an intrinsic active ringdown quencher, thus other active quenchers suggested for high-Q tanks (see e.g. [39],[31]) were considered as unnecessary. They could, of course, still be added in order to suppress residual leakage of low frequency ringing of the biasing network due to parasitic switching spikes, if necessary.

In our setup probeheads A and B yield reasonably small ringdown artifacts (<50nV) with an additional blanking time $T_B$ of 4µs after $T_W$ which allows for the acquisition of fast decaying FID signals. Extending $T_B$ to 10µs eliminated the artefacts below detectability when averaging up to N=2000. Probehead C, as expected, requires somewhat longer blanking times. When carrying out narrowband spectroscopy with a commercial spectrometer and a commercial, motor-tuned classical probehead (Tabletop NQR Probe Autotune, NMR Service GmbH, Germany) the required blanking time was approximately the same, thus no significant difference could be found between both systems.

Further improvements could probably be achieved with more elaborated pulse techniques, or the subtraction of two immediately subsequent measurements with and without saturation of the spin population ([40], [41]). The latter technique, however, is only applicable if $T_1$ is significantly longer than the interpulse interval. The use of short counterphase pulses after the main RF pulse may also be considered [42], however, this method is rather complex and wideband operation may become a major challenge.

A last point is the length of the TX pulse. While with classical high-Q-resonators we can use short pulses down to 1 µs for a 90° excitation at moderate powers (100 – 500W), the wideband probehead requires significantly longer pulses for the same flip angle because of the low Q during TX. Though this problem decreases with decreasing frequency (higher number of turns and thus higher flux density), it is intrinsic in the design and limits the detectability of peaks with short $T_2$.

We shall also mention two additional, but minor caveats:

(1) Our stripline design was designed on a heuristic basis rather than using more extended simulations with specialized software. Thus there may still be room for a slight improvement of the cryo-nozzle.

(2) The NWA-based spectrometer does not allow for more elaborate pulse sequences supporting e.g. phase cycling. Therefore accurate quantitative measurements e.g. of the relaxation time constants like in [43] are not possible.

To summarize, our approach can compete in all important aspects with the classical one while it can provide significant acceleration of the total measuring time when searching for unknown NQR transitions of a newly synthetized compound. Especially for transitions with long $T_1$ the approach is extremely useful and the practical operation is usually less complicated than with motor-driven probeheads, which require complex control mechanisms.

## Acknowledgements

This project has received funding from the European Union's *Horizon 2020 research and innovation programme* under grant agreement No 665172